\documentclass[runningheads]{llncs}
\usepackage{graphicx}
\usepackage{todonotes}

\usepackage{multirow}
\usepackage[table,xcdraw]{}
\usepackage[normalem]{ulem}
\useunder{\uline}{\ul}{}

\begin{document}
\title{Shifted Windows Transformers for Medical Image Quality Assessment}

%
%
\author{Caner Özer\inst{*, 1} \and
Arda Güler\inst{2} \and
Aysel Türkvatan Cansever\inst{2} \and
Deniz Alis\inst{3} \and
Ercan Karaarslan\inst{3} \and
İlkay Öksüz\inst{1,4} 
}
\authorrunning{Ozer et. al.}
%

\institute{Istanbul Technical University, Istanbul, Turkey \and
Mehmet Akif Ersoy Thoracic and Cardiovascular Surgery Training and Research Hospital, Istanbul, Turkey \and
Acibadem University, Istanbul, Turkey \and
King's College London\\
$^*$Corresponding author: \email{ozerc@itu.edu.tr}
}
\maketitle              
\begin{abstract}
To maintain a standard in a medical imaging study, images should have necessary image quality for potential diagnostic use. Although CNN-based approaches are used to assess the image quality, their performance can still be improved in terms of accuracy. In this work, we approach this problem by using Swin Transformer, which improves the poor-quality image classification performance that causes the degradation in medical image quality. We test our approach on Foreign Object Classification problem on Chest X-Rays (Object-CXR) and Left Ventricular Outflow Tract Classification problem on Cardiac MRI with a four-chamber view (LVOT). While we obtain a classification accuracy of 87.1\% and 95.48\% on the Object-CXR and LVOT datasets, our experimental results suggest that the use of Swin Transformer improves the Object-CXR classification performance while obtaining a comparable performance for the LVOT dataset. To the best of our knowledge, our study is the first vision transformer application for medical image quality assessment.

\end{abstract}

\section{Introduction}

It is the responsibility of technicians to ensure that the acquired medical scan should be of good quality so that the physician, who asked for the medical scan, will be able to diagnose accurately. However, this condition may not be satisfied due to the medical image quality problems such as patient breathing or arrhythmia in Cardiac Magnetic Resonance Imaging (MRI), ringing artifact presence in Computed Tomography (CT), and low dose or foreign object appearance on the Chest X-Rays (CXR). These image quality issues would also constitute a critical problem for automated processing of medical scans within a clinical decision support system consisting of big data. 


Maintaining a high standard in image quality is imperative also for a machine learning-guided medical image analysis downstream tasks. More specifically, existence of poor quality images may lead a machine learning methodology to learn spurious set of features which are not clinically useful. Hence, identifying, eliminating or correcting the samples with poor image quality is necessary for downstream tasks such as segmentation and disease diagnosis. The reasons for an image that should be eliminated from the pipeline include but are not limited to artefacts in MR Imaging \cite{Ma2020}, foreign object existence \cite{Xue2015}, or undesirable region appearance \cite{Oksuz2018} within the scans. For this purpose, several datasets on a variety of modalities have been developed to overcome such problems. Ma et. al. \cite{Ma2020} has developed a brain MRI dataset in which the patient data has been labelled according to the level of motion artefact. The authors have also developed a 4-layer convolutional neural network model to classify the samples according to their motion degradation level. JFHealthcare \cite{CXR20} proposed a MIDL 2020 challenge which comes up with a dataset that aims to localize the foreign objects within chest X-Rays.

Swin-Transformers are becoming more common in medical image analysis, especially in the field of diagnosis. For instance, Zhang et al. \cite{Zhang2021} uses a cascaded mechanism for chest CT COVID-19 diagnosis which uses a U-Net to extract lungs as a foreground and background segmentation mechanism, and Swin-Transformer uses these segmented lungs to find out whether the patient has COVID-19 or not. Chen et al. \cite{Chen2021} combines a statistical significance-based slice sampling scheme and Swin-Transformers to propose a framework that can detect COVID-19 from chest CT scans by using a single slice. In addition to diagnosis, Swin-Transformers are becoming more common to be used for other problems, such as brain tumor segmentation \cite{Cao2021} \cite{Hatamizadeh2022}, and cardiac CT reconstruction \cite{Pan2021}. However, despite being promising for these problems in medical imaging, Swin-Transformers' have yet to be studied for medical image quality assessment tasks.

In this work, we propose a SwinTransformer-based \cite{Liu2021} approach for medical image quality analysis where our contributions are as follows:

\begin{itemize}
    \item We train a variety of SwinTransformer models in order to classify X-Ray images with foreign objects and Cardiac MRI images with a 4-chamber view with Left Ventricular Outflow Tract (LVOT).
    \item We compare our approach with deep neural network (DNN) based approaches, such as ResNet \cite{He2016} and EfficientNet \cite{Tan2019}.
    \item To our best knowledge, this is the first paper to perform medical image quality assessment through SwinTransformers. 

\end{itemize}

\section{Methodology}


\begin{figure}
    \centering
    \includegraphics[width=\textwidth]{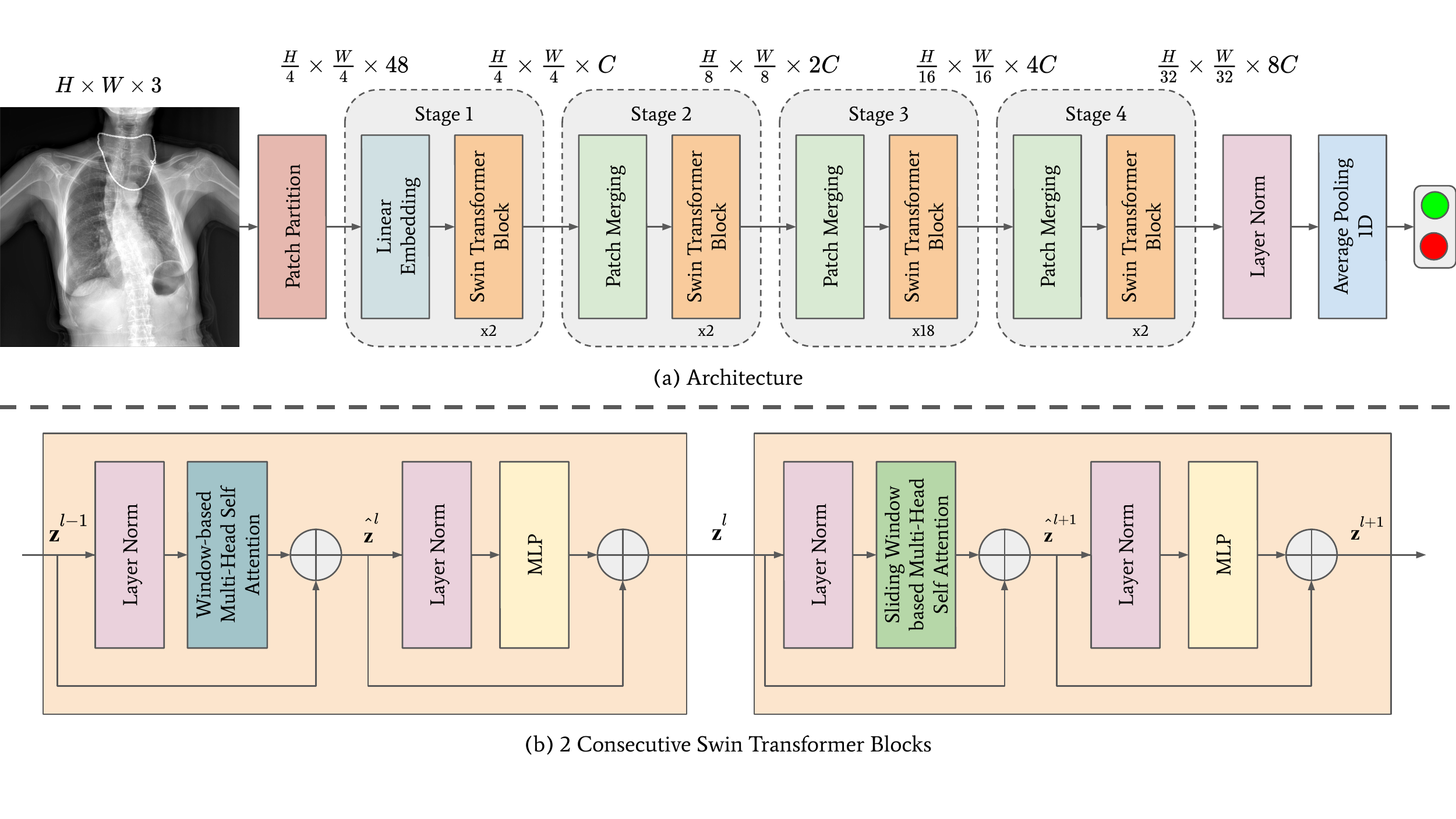}
    \caption{Swin Transformer pipeline for medical image quality analysis.}
    \label{fig:swin_visual}
\end{figure}

\subsection{Pre-processing}
In order to train our model, we apply a number of preprocessing procedures prior to feeding the input images to the network. In this regard, instead of manually defining the data augmentations and their corresponding hyperparameters, we use RandAugment \cite{Cubuk2020} which uses a previously found automated data augmentation strategy to train a DNN. After that, we resize these images to a pre-determined resolution of $H \times W$. Finally, we normalize the images by using ImageNet statistics that they become ready-to-use in a pre-trained model. 

\subsection{Swin Transformer}
Following the pre-processing procedure, we used the Swin Transformer model that is being demonstrated in Fig. \ref{fig:swin_visual}. Given an input image, Swin Transformer first divides the image into patches using a Patch Partition layer that generates $48$-dimensional representations for each of the $4 \times 4$ patches on the image. After that, as shown in Stage 1, the Linear Embedding layer transforms $48$-d representations into $C$ dimensions.

The second component of Stage 1 is the Swin Transformer Blocks which has been inherited from the general Visual Transformer (ViT) \cite{Dosovitskiy2021} architecture. Despite having the same high-level block structure as the one in ViT, multiheaded self-attention (MSA) blocks are replaced with an alternate ordering of window-based MSA (W-MSA) and shifted window-based MSA (SW-MSA), sequentially. In W-MSA, the feature map is divided into windows and MSA computes the output on each of them. However, this brings a cost of missing global connectivity, since this was not an issue for MSA in ViT that operates globally. For this reason, cyclic shifting operation is used to overcome this problem with SW-MSA that, the feature map is circularly shifted on the height and width axes by the half-length of the patch size. Then, W-MSA is executed on the resulting feature map, and a reverse circular shift operation is used to obtain a reverted feature map without any shifts. 

In all of Stages 2, 3, and 4, we have a Patch Merging layer and a varying number of Swin Transformer blocks. In the Patch Merging layer, each of the pixels located on the $2 \times 2$ patches on the feature map are concatenated across the channel axis, i.e., $\frac{H}{2} \times \frac{W}{2} \times C$ to $\frac{H}{4} \times \frac{W}{4} \times 4C$. Then, after applying the layer normalization layer, a linear layer is used to reduce the dimensionality to half, i.e., to $\frac{H}{4} \times \frac{W}{4} \times 2C$. The output is then transferred to the Swin Transformer blocks for further processing. Lastly, after using layer normalization followed by an average pooling layer to calculate the features of the Swin Transformer model, the output is used for classification in a single Linear layer. This model classifies if there is anything present in the input image that can affect the image quality adversely. The hyperparameter settings can be found in Table \ref{tab:swint_hyperparams}. Stochastic depth refers to the probability of disregarding the transformation that is applied on a sample within a batch during training, while being processed within a Swin Transformer block, embedding dims $C$ is the dimensionality of the patch-wise embeddings, depths and MSA heads correspond to the number of repeated Swin Transformer blocks and head count of the MSA layer for each stage, respectively.

\begin{table}[]
\centering
\caption{Hyperparameter settings for different Swin Transformer models.}
\begin{tabular}{|l|c|c|c|c|}
\hline
Model Name & \begin{tabular}[c]{@{}c@{}}Stochastic Depth\end{tabular} & \begin{tabular}[c]{@{}c@{}}Embedding Dims $(C)$\end{tabular} & Depths      & MSA Heads    \\ \hline
Swin-Tiny  & 0.2                                                        & 96                                                             & 2, 2, 6, 2  & 3, 6, 12, 24 \\ \hline
Swin-Small & 0.3                                                        & 96                                                             & 2, 2, 18, 2 & 3, 6, 12, 24 \\ \hline
Swin-Base  & 0.2                                                        & 128                                                            & 2, 2, 18, 2 & 4, 8, 16, 32 \\ \hline
\end{tabular}
\label{tab:swint_hyperparams}
\end{table}

\subsection{Training the Model}
We adopt the official code of Swin Transformers\footnotemark{} in which we performed some modifications to run it on PyTorch 1.10.2. We also used TorchVision 0.11.3 to inherit the pre-trained models for EfficientNet and ResNet pre-trained architectures. We apply the learning rate warm-up scheme and learning rate scheduling and leverage color jittering, stochastic depth (only for Swin Transformer) \cite{Huang2016}, Random Erasing \cite{Zhong2017} MixUp \cite{Zhang2018}, CutMix \cite{Yun2019} and RandAugment \cite{Cubuk2020} data augmentation methods for promoting model generalization. For the foreign object classification model, we set the batch size to 16, gradient accumulation steps to 8 and window size to 8. Instead for the LVOT classification model, we set the batch size to 64, gradient accumulation steps to 2 and window size to 7, due to lower input image resolution comparing chest X-Rays. We employ an AdamW \cite{Kingma2015} optimizer with weight-decay of $10^{-8}$ to fine-tune both of these models by using an ImageNet pretrained model, as we set the initial learning rate to $0.00006$ and use a linear learning-rate scheduler. The fine-tuning procedure takes $60$ epochs with $5$ epochs of a warm-up period. We also trained the models from scratch for $300$ epochs that, the learning rate is set to $0.0005$ and the objective function is Cross Entropy. You can access to our repository by using the following link: \url{https://github.com/canerozer/qct}. \footnotetext{https://github.com/microsoft/Swin-Transformer}

\section{Experimental Results}

\subsection{Datasets}

We demonstrate our results on the Object-CXR \cite{CXR20}, a challenge dataset where the aim is to classify the Chest X-Ray scans with foreign objects and to localize the foreign objects, and the LVOT classification dataset, which is constructed by selecting a subset of the Cardiac MR (CMR) scans with 4-chamber view that were acquired in the Mehmet Akif Ersoy Thoracic and Cardiovascular Surgery Training and Research Hospital between the years 2016 and 2019. Object-CXR dataset contains 8,000 samples for training and 1,000 samples for each of the validation and testing splits. There is an equal number of positive and negative samples for each split which does and does not contain foreign objects. We only use the class label, but not the localization ground truth for this dataset during training. We set the fixed input size to $H = W = 1024$ since the images can reach a spatial resolution of $4048 \times 4932$. On the other hand, LVOT classification dataset contains 272 positive and 278 negative samples that are used for training, 34 positive and 35 negative samples for validation, and 35 positive and 35 negative samples for testing. These images have relatively way less spatial resolution than the ones in Object-CXR, hence, we set $H = W = 224$. In addition, since our pipeline is capable of processing 2D images, we slice the 3D CMR patient scans into 2D images, which is also helpful in terms of increasing the number of samples for each split by a factor of 25.

\subsection{Left Ventricular Outflow Tract Classification Results}

First, we start by comparing the performances of different Swin Transformer models that are named as Swin-Tiny, Swin-Small, and Swin-Base in Table \ref{tab:lvot:swint}. We additionally measured the differences between using a pretrained model and training from scratch. 
We notice that, the accuracy is the highest for the Swin-Tiny model with an accuracy of $95.59\%$. Nevertheless, Swin-Small's AUC (Area Under Curve) score is performing the best among these three models with a score of $0.980$. For this reason, we decided to include both of the Swin-Small and Swin-Tiny models to compare them with the other baselines. Meanwhile, switching from pre-trained model to training from scratch significantly decreases the accuracy of at least around $12.35\%$ and AUC score of $0.059$. We relate this to the data-hungry regime of Swin Transformers, as they need abundant data samples to capture the inductive bias during training.
Hence, it becomes necessary to use the pretrained models or to increase the number of samples to the order of tens of thousands.

\begin{table}[h!]
\caption{Comparison among Swin Transformer models with different capacity for the LVOT classification task. }
\centering
\begin{tabular}{|l|c|c|c|c|c|c|}
\hline
Model                       & \multicolumn{1}{c|}{Pretrained?} & \multicolumn{1}{c|}{Acc} & \multicolumn{1}{c|}{AUC} & \multicolumn{1}{c|}{\begin{tabular}[c]{@{}c@{}}\# of params\end{tabular}} & \multicolumn{1}{c|}{GFLOPs} & \multicolumn{1}{c|}{\begin{tabular}[c]{@{}c@{}}Throughput (img/sec)\end{tabular}} \\ \hline
\multirow{2}{*}{Swin-Tiny}  & Yes                              & {\ul \textbf{95.59}}     & 0.969                    & \multirow{2}{*}{27,520,892}                                                 & \multirow{2}{*}{4.5}        & \multirow{2}{*}{515}                                                                \\ \cline{2-4}
                            & No                               & 72.93                    & 0.757                    &                                                                             &                             &                                                                                     \\ \hline
\multirow{2}{*}{Swin-Small} & Yes                              & 95.48                    & {\ul \textbf{0.980}}     & \multirow{2}{*}{48,838,796}                                                 & \multirow{2}{*}{8.7}        & \multirow{2}{*}{313}                                                                \\ \cline{2-4}
                            & No                               & 83.13                    & 0.921                    &                                                                             &                             &                                                                                     \\ \hline
\multirow{2}{*}{Swin-Base}  & Yes                              & 93.86                    & 0.976                    & \multirow{2}{*}{86,745,274}                                                 & \multirow{2}{*}{15.4}       & \multirow{2}{*}{222}                                                                \\ \cline{2-4}
                            & No                               & 67.01                    & 0.738                    &                                                                             &                             &                                                                                     \\ \hline
\end{tabular}

\label{tab:lvot:swint}
\end{table}

\begin{table}[]
\caption{Our results on the LVOT dataset indicate that the performance is comparable to the other DNN-based methods.}
\centering
\begin{tabular}{|l|c|c|c|c|c|}
\hline
Model           & Acc                  & AUC                  & \begin{tabular}[c]{@{}c@{}}\# of params\end{tabular} & GFLOPs & Throughput \\ \hline
EfficientNet-B0 & 93.74                & 0.979                & 4,010,110                                              & 0.4    & 2274       \\
EfficientNet-B1 & 93.68                & 0.971                & 6,515,746                                              & 0.6    & 1565       \\
EfficientNet-B2 & 94.61                & 0.971                & 7,703,812                                              & 0.7    & 1480       \\
ResNet-34       & 95.19                & 0.974                & 21,285,698                                             & 3.7    & 1963       \\
ResNet-50       & 95.77                & 0.979                & 23,512,130                                             & 4.1    & 1005       \\
ResNet-101      & 95.59                & 0.969                & 42,504,258                                             & 7.8    & 587        \\
ResNet-152      & {\ul \textbf{96.00}} & {\ul \textbf{0.988}} & 58,147,906                                             & 11.6   & 409        \\ \hline \hline
Swin-Tiny       & 95.59                & 0.969                & 27,520,892                                             & 4.5    & 515        \\
Swin-Small      & 95.48                & 0.980                & 48,838,796                                             & 8.7    & 313        \\ \hline
\end{tabular}
\label{tab:lvot:all}
\end{table}

Then, we compare our approach with other deep learning baselines as ResNet \cite{He2016} and EfficientNet \cite{Tan2019} as shown in Table \ref{tab:lvot:all}. We see that, despite the differences in the number of parameters and GFLOPs, the performance varies insignificantly ($p > 0.05$) with outstanding accuracy and AUC metrics. We relate this to the triviality of the LVOT classification task, where the goal is to check if fifth chamber exists on the Cardiac MRI scan - generally located at the centre. As a result, it becomes possible to obtain testing accuracy of over $95\%$ and AUC scores exceeding $0.969$ for all ResNet architectures and Swin-Small without overfitting. Still, these models are ready to be deployed for a real-world scenario for alerting LVOT regions on 4-chamber Cardiac MRIs almost instantaneously.  

\subsection{Foreign Object Classification Results}

\begin{table}[]
\caption{Results for the Object-CXR dataset.}
\centering
\begin{tabular}{|l|c|c|c|c|c|}
\hline
Model           & Acc                 & AUC                  & \begin{tabular}[c]{@{}c@{}}\# of params\end{tabular} & GFLOPs & Throughput \\ \hline
EfficientNet-B0 & 82.1                & 0.873                & 4,010,110                                              & 8.3    & 119        \\
EfficientNet-B1 & 83.0                & 0.882                & 6,515,746                                              & 12.3   & 83         \\
EfficientNet-B2 & 83.2                & 0.890                & 7,703,812                                              & 14.2   & 78         \\
ResNet-34       & 82.9                & 0.895                & 21,285,698                                             & 91.8   & 94         \\
ResNet-50       & 82.8                & 0.897                & 23,512,130                                             & 102.7  & 42         \\
ResNet-101      & 84.1                & 0.906                & 42,504,258                                             & 195.8  & 26         \\
ResNet-152      & 85.1                & 0.909                & 58,147,906                                             & 241.5  & 22         \\ \hline \hline
Swin-Base       & {\ul \textbf{87.1}} & {\ul \textbf{0.922}} & 86,766,330                                             & 324.6  & 11         \\ \hline
\end{tabular}
\label{tab:ocxr}
\end{table}

The results show Swin Transformer's strength on a higher dimensional input, namely, on the Chest X-Ray data for foreign object classification. The model complexity is highly correlated with the model performance that, the best performance is obtained by the Swin-Base model with $87.1\%$ testing accuracy and $0.922$ testing AUC score. The closest performance to the best model is the ResNet-152 model, which has $1.5 \times$ less number of parameters and $25\%$ less floating point operations, at a cost of $2\%$ less testing accuracy and $0.013$ less AUC score. Despite Swin-Base is working slightly slower than real-time due to its huge input size, its significant performance gains ($p < 0.005$) makes it a promising approach. 

\subsection{Qualitative Analysis}

\begin{figure}
    \centering
    \includegraphics[width=\textwidth]  {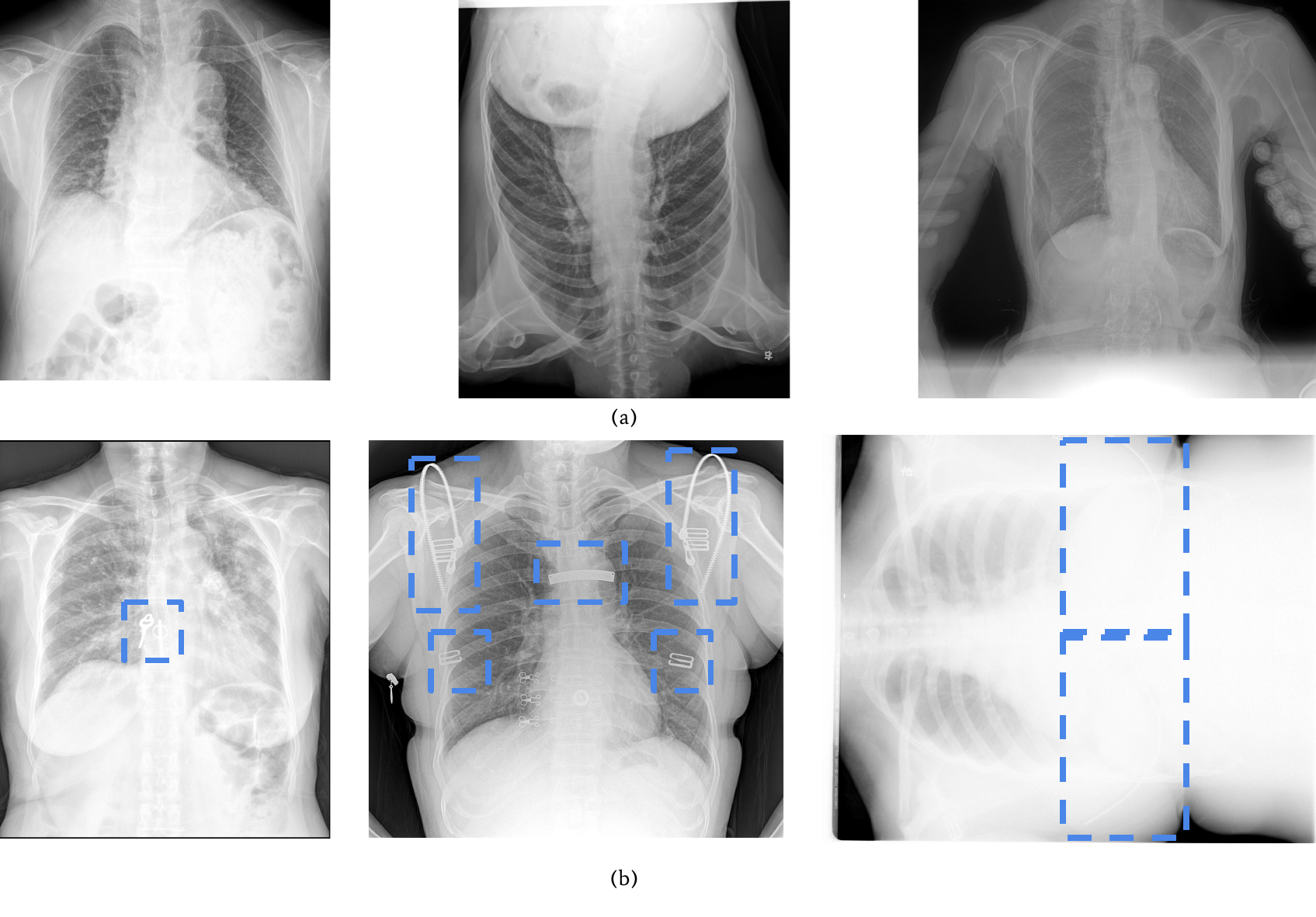}
    \caption{Some true challenging samples due to their medical image quality factors, from Object-CXR dataset.}
    \label{fig:ocxr:hard}
\end{figure}

Prior to utilizing our model in a clinical study, we further examine our model under extreme conditions. Firstly, we start with the Object-CXR classification task as some cases are shown in Fig. \ref{fig:ocxr:hard}. Regardless of the orientation and contrast factors, our model can correctly classify the X-Ray images which does and does not contain foreign objects as demonstrated in Fig. \ref{fig:ocxr:hard}a and \ref{fig:ocxr:hard}b, respectively. What is interesting especially for the rightmost image of Fig. \ref{fig:ocxr:hard}b is that, even though the foreign objects may not be visible to human eye without altering the contrast, our model can still successfully classify the image as it contains foreign objects. Foreign objects are marked with a blue bounding box in Fig. \ref{fig:ocxr:hard}b.

\begin{figure}
    \centering
    \includegraphics[width=\textwidth]  {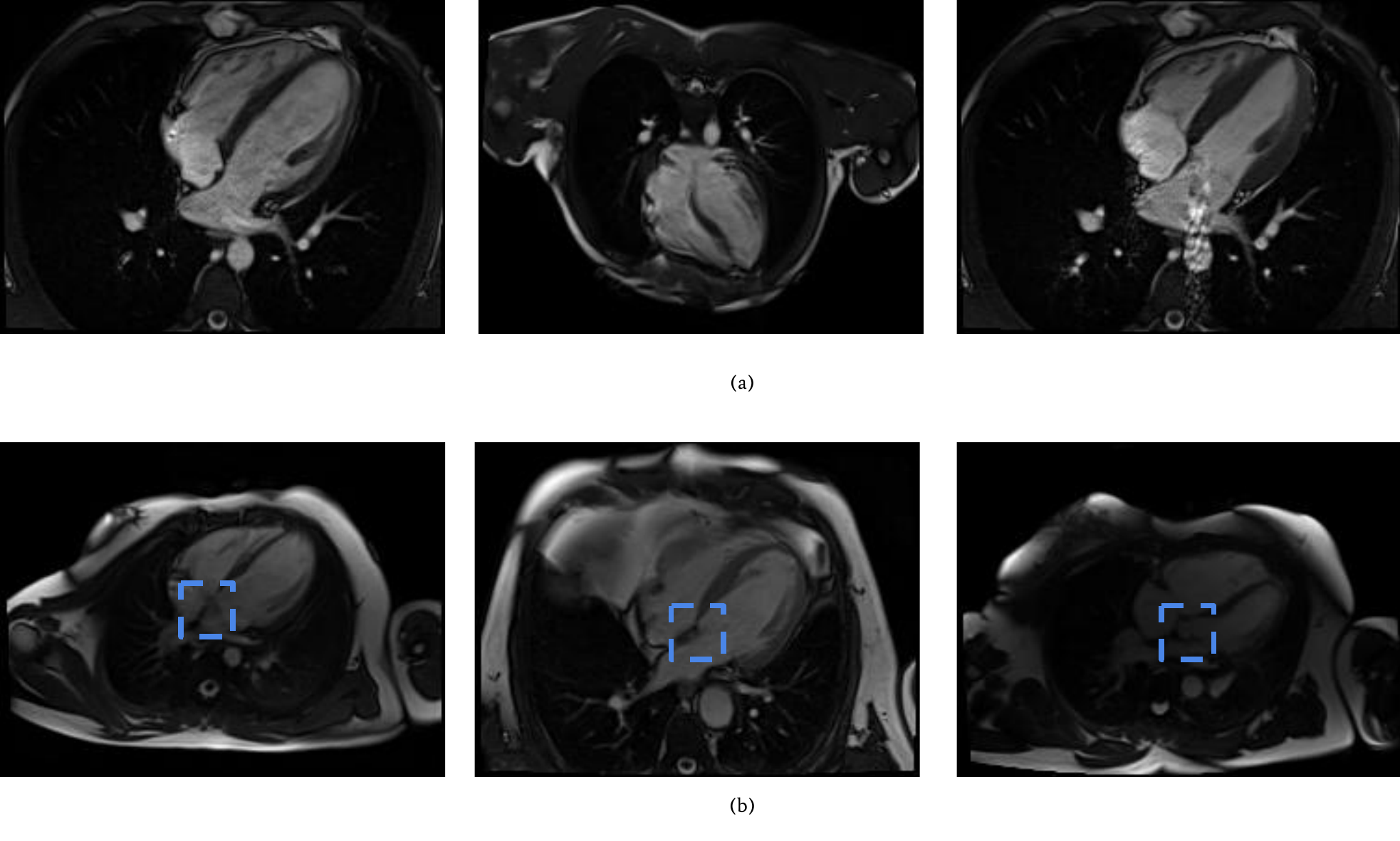}
    \caption{Some true challenging samples due to their medical image quality factors, from LVOT dataset.}
    \label{fig:lvot:hard}
\end{figure}

We also perform a qualitative analysis on the classification results as some of the edge cases are shown in Fig. \ref{fig:lvot:hard} for the LVOT classification task. In Fig. \ref{fig:lvot:hard}a, where all 3 images are labelled as good quality images without LVOT appearance, the model can correctly classify even when the images are subject to motion artefacts and local noise upto some level. Also, as shown in Fig. \ref{fig:lvot:hard}b, our model can correctly classify the images containing LVOT regions under low contrast. We show the regions with LVOT with a blue bounding box in Fig. \ref{fig:lvot:hard}b. 

\section{Discussion \& Conclusion}
In this work, we propose to use Swin Transformers for medical image quality analysis, where we validated our approach on two datasets working on two different modalities. We outperform a variety of ResNet and EfficientNet baselines on the Object-CXR dataset by obtaining a testing accuracy of $87.1\%$ while obtaining a comparable performance for the LVOT classification objective. Comparing Swin Transformer to the other baselines and performing a qualitative analysis demonstrate its potential for utilizing it in a clinical setting. 

Although the problem setting has been restricted to foreign object classification in Chest X-Rays and LVOT classification for Cardiac MRI's, it is possible to generalize the approach for other medical image quality problems such as motion artefact grade classification. 
In addition, the interpretability aspect of the Swin Transformers is a critical avenue of research for the safe transition of an automated medical image analysis pipeline.

\section*{Acknowledgments}
This paper has been produced benefiting from the 2232 International Fellowship for Outstanding Researchers Program of TUBITAK (Project No: 118C353). However, the entire responsibility of the publication/paper belongs to the owner of the paper. The financial support received from TUBITAK does not mean that the content of the publication is approved in a scientific sense by TUBITAK.



%
%
%
\bibliographystyle{splncs04}
\bibliography{mybibliography}

\begin{thebibliography}{10}
\providecommand{\url}[1]{\texttt{#1}}
\providecommand{\urlprefix}{URL }
\providecommand{\doi}[1]{https://doi.org/#1}

\bibitem{Cao2021}
Cao, H., Wang, Y., Chen, J., Jiang, D., Zhang, X., Tian, Q., Wang, M.:
  Swin-unet: Unet-like pure transformer for medical image segmentation (2021)

\bibitem{Chen2021}
Chen, G.L., Hsu, C.C., Wu, M.H.: Adaptive distribution learning with
  statistical hypothesis testing for covid-19 ct scan classification. In: 2021
  IEEE/CVF International Conference on Computer Vision Workshops (ICCVW). pp.
  471--479 (2021). \doi{10.1109/ICCVW54120.2021.00057}

\bibitem{Cubuk2020}
Cubuk, E.D., Zoph, B., Shlens, J., Le, Q.: Randaugment: Practical automated
  data augmentation with a reduced search space. In: Larochelle, H., Ranzato,
  M., Hadsell, R., Balcan, M.F., Lin, H. (eds.) Advances in Neural Information
  Processing Systems. vol.~33, pp. 18613--18624. Curran Associates, Inc.
  (2020),
  \url{https://proceedings.neurips.cc/paper/2020/file/d85b63ef0ccb114d0a3bb7b7d808028f-Paper.pdf}

\bibitem{Dosovitskiy2021}
Dosovitskiy, A., Beyer, L., Kolesnikov, A., Weissenborn, D., Zhai, X.,
  Unterthiner, T., Dehghani, M., Minderer, M., Heigold, G., Gelly, S.,
  Uszkoreit, J., Houlsby, N.: An image is worth 16x16 words: Transformers for
  image recognition at scale. In: 9th International Conference on Learning
  Representations, {ICLR} 2021, Virtual Event, Austria, May 3-7, 2021.
  OpenReview.net (2021), \url{https://openreview.net/forum?id=YicbFdNTTy}

\bibitem{Hatamizadeh2022}
Hatamizadeh, A., Nath, V., Tang, Y., Yang, D., Roth, H., Xu, D.: Swin {UNETR:}
  swin transformers for semantic segmentation of brain tumors in {MRI} images.
  CoRR  \textbf{abs/2201.01266} (2022), \url{https://arxiv.org/abs/2201.01266}

\bibitem{He2016}
He, K., Zhang, X., Ren, S., Sun, J.: Deep residual learning for image
  recognition. In: 2016 {IEEE} Conference on Computer Vision and Pattern
  Recognition, {CVPR} 2016, Las Vegas, NV, USA, June 27-30, 2016. pp. 770--778.
  {IEEE} Computer Society (2016). \doi{10.1109/CVPR.2016.90},
  \url{https://doi.org/10.1109/CVPR.2016.90}

\bibitem{Huang2016}
Huang, G., Sun, Y., Liu, Z., Sedra, D., Weinberger, K.Q.: Deep networks with
  stochastic depth. In: Leibe, B., Matas, J., Sebe, N., Welling, M. (eds.)
  Computer Vision - {ECCV} 2016 - 14th European Conference, Amsterdam, The
  Netherlands, October 11-14, 2016, Proceedings, Part {IV}. Lecture Notes in
  Computer Science, vol.~9908, pp. 646--661. Springer (2016).
  \doi{10.1007/978-3-319-46493-0\_39},
  \url{https://doi.org/10.1007/978-3-319-46493-0\_39}

\bibitem{CXR20}
JFHealthcare: Automatic detection of foreign objects on chest x-rays.
  \url{https://academictorrents.com/details/fdc91f11d7010f7259a05403fc9d00079a09f5d5}
  (2020), accessed on 11/27/2020

\bibitem{Kingma2015}
Kingma, D.P., Ba, J.: Adam: {A} method for stochastic optimization. In: Bengio,
  Y., LeCun, Y. (eds.) 3rd International Conference on Learning
  Representations, {ICLR} 2015, San Diego, CA, USA, May 7-9, 2015, Conference
  Track Proceedings (2015), \url{http://arxiv.org/abs/1412.6980}

\bibitem{Liu2021}
Liu, Z., Lin, Y., Cao, Y., Hu, H., Wei, Y., Zhang, Z., Lin, S., Guo, B.: Swin
  transformer: Hierarchical vision transformer using shifted windows. In:
  Proceedings of the IEEE/CVF International Conference on Computer Vision
  (ICCV). pp. 10012--10022 (October 2021)

\bibitem{Ma2020}
Ma, J.J., Nakarmi, U., Kin, C.Y.S., Sandino, C.M., Cheng, J.Y., Syed, A.B.,
  Wei, P., Pauly, J.M., Vasanawala, S.S.: Diagnostic image quality assessment
  and classification in medical imaging: Opportunities and challenges. In: 17th
  {IEEE} International Symposium on Biomedical Imaging, {ISBI} 2020, Iowa City,
  IA, USA, April 3-7, 2020. pp. 337--340. {IEEE} (2020).
  \doi{10.1109/ISBI45749.2020.9098735},
  \url{https://doi.org/10.1109/ISBI45749.2020.9098735}

\bibitem{Oksuz2018}
Oksuz, I., Ruijsink, B., Puyol-Antón, E., Sinclair, M., Rueckert, D.,
  Schnabel, J.A., King, A.P.: Automatic left ventricular outflow tract
  classification for accurate cardiac mr planning. In: 2018 IEEE 15th
  International Symposium on Biomedical Imaging (ISBI 2018). pp. 462--465
  (2018). \doi{10.1109/ISBI.2018.8363616}

\bibitem{Pan2021}
Pan, J., Wu, W., Gao, Z., Zhang, H.: Mist-net: Multi-domain integrative swin
  transformer network for sparse-view {CT} reconstruction. CoRR
  \textbf{abs/2111.14831} (2021), \url{https://arxiv.org/abs/2111.14831}

\bibitem{Tan2019}
Tan, M., Le, Q.V.: Efficientnet: Rethinking model scaling for convolutional
  neural networks. In: Chaudhuri, K., Salakhutdinov, R. (eds.) Proceedings of
  the 36th International Conference on Machine Learning, {ICML} 2019, 9-15 June
  2019, Long Beach, California, {USA}. Proceedings of Machine Learning
  Research, vol.~97, pp. 6105--6114. {PMLR} (2019),
  \url{http://proceedings.mlr.press/v97/tan19a.html}

\bibitem{Xue2015}
Xue, Z., Candemir, S., Antani, S., Long, L.R., Jaeger, S., Demner-Fushman, D.,
  Thoma, G.R.: Foreign object detection in chest x-rays. In: 2015 IEEE
  International Conference on Bioinformatics and Biomedicine (BIBM). pp.
  956--961 (2015). \doi{10.1109/BIBM.2015.7359812}

\bibitem{Yun2019}
Yun, S., Han, D., Chun, S., Oh, S.J., Yoo, Y., Choe, J.: Cutmix: Regularization
  strategy to train strong classifiers with localizable features. In: 2019
  {IEEE/CVF} International Conference on Computer Vision, {ICCV} 2019, Seoul,
  Korea (South), October 27 - November 2, 2019. pp. 6022--6031. {IEEE} (2019).
  \doi{10.1109/ICCV.2019.00612}, \url{https://doi.org/10.1109/ICCV.2019.00612}

\bibitem{Zhang2018}
Zhang, H., Ciss{\'{e}}, M., Dauphin, Y.N., Lopez{-}Paz, D.: mixup: Beyond
  empirical risk minimization. In: 6th International Conference on Learning
  Representations, {ICLR} 2018, Vancouver, BC, Canada, April 30 - May 3, 2018,
  Conference Track Proceedings. OpenReview.net (2018),
  \url{https://openreview.net/forum?id=r1Ddp1-Rb}

\bibitem{Zhang2021}
Zhang, L., Wen, Y.: A transformer-based framework for automatic covid19
  diagnosis in chest cts. In: 2021 IEEE/CVF International Conference on
  Computer Vision Workshops (ICCVW). pp. 513--518 (2021).
  \doi{10.1109/ICCVW54120.2021.00063}

\bibitem{Zhong2017}
Zhong, Z., Zheng, L., Kang, G., Li, S., Yang, Y.: Random erasing data
  augmentation. In: The Thirty-Fourth {AAAI} Conference on Artificial
  Intelligence, {AAAI} 2020, The Thirty-Second Innovative Applications of
  Artificial Intelligence Conference, {IAAI} 2020, The Tenth {AAAI} Symposium
  on Educational Advances in Artificial Intelligence, {EAAI} 2020, New York,
  NY, USA, February 7-12, 2020. pp. 13001--13008. {AAAI} Press (2020),
  \url{https://aaai.org/ojs/index.php/AAAI/article/view/7000}

\end{thebibliography}
%




\end{document}


\title{Shifted Windows Transformers for Medical Image Quality Assessment}
%
%
\author{Anonymous Authors\inst{1}}
%
\authorrunning{Anonymous}
%
%
\maketitle              

\section{False Analysis}

\begin{figure}
    \centering
    \includegraphics[width=0.8\textwidth]{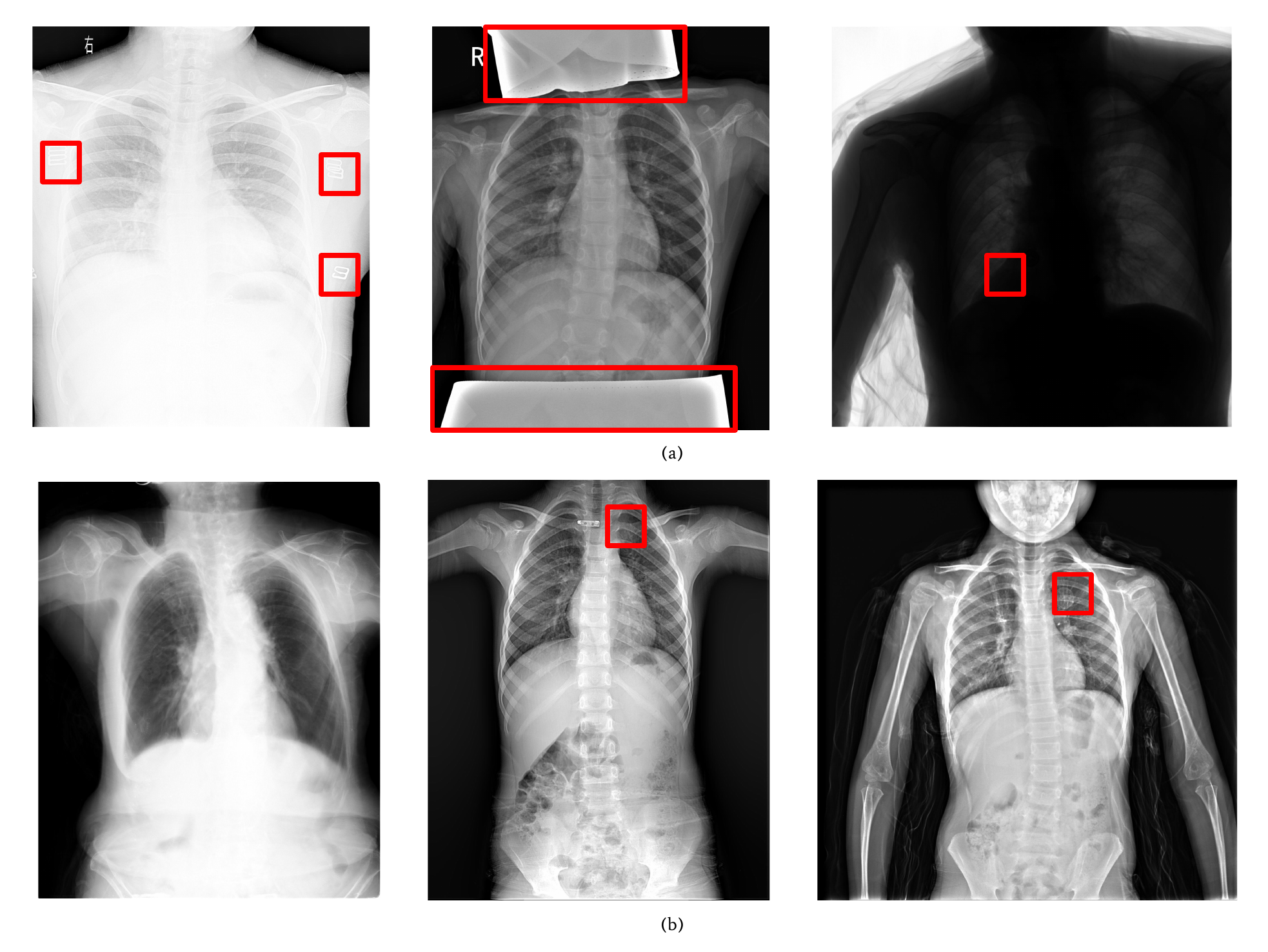}
    \caption{Some samples regarding the false positive (a) and false negative (b) classifications of the Swin-Base model on the Object-CXR dataset. (Best viewed in zoom)}
    \label{fig:cxr:fp_fn}
\end{figure}

We also perform a qualitative analysis on the classification results on the edge cases as some of them are shown in Fig. \ref{fig:cxr:fp_fn}. In Fig. \ref{fig:cxr:fp_fn}a, the model classifies all of these 3 images as images with foreign objects, however, ground-truth says the reverse. Nevertheless, in our qualitative study, we noticed that some foreign objects are clearly visible, even though the images are marked as false positives. We highlight the visible foreign objects with a red bounding box in Fig. \ref{fig:cxr:fp_fn}, where we needed to use histogram equalization on the right-most image to see the button. For the false negatives as the ones in Fig. \ref{fig:cxr:fp_fn}b, we identified that our model may not classify the foreign objects when there exists motion blur or the x-ray covers an area beyond the lungs. It is possible to alleviate this problem by extracting the pulmonary region from the chest X-Rays to apply a constraint on the area of interest. Except for the image at the top-right, which is classified as a good quality image with high entropy value, all of these edge cases are also valid for the second best DNN model, ResNet-152.

\begin{figure}
    \centering
    \includegraphics[width=\textwidth]  {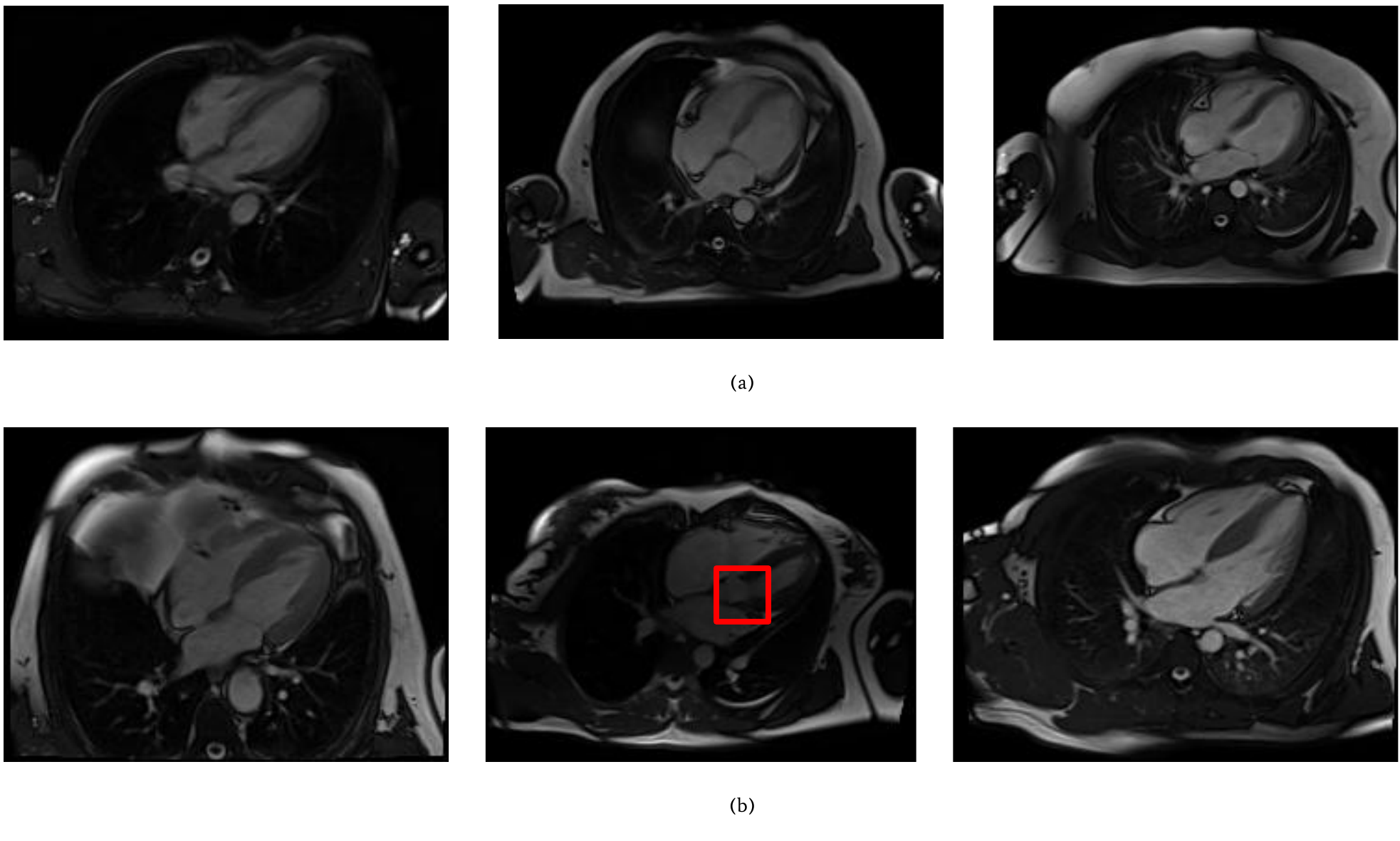}
    \caption{Some false positive (a) and false negative (b) samples as a result of using the classification output of Swin-Small model on the LVOT dataset. (Best viewed in zoom)}
    \label{fig:lvot:fp_fn}
\end{figure}

We also perform the same analysis on the LVOT classification dataset, where some false positive and negatives are shown in Fig. \ref{fig:lvot:fp_fn}. We notice that motion blur and contrast are the primary causes of model performance degradation such that identifying the image quality in those aspects are more critical to proceed with the following procedures of a medical image analysis pipeline. All of these edge cases are also misclassified for the best model, ResNet-152. 

